\documentclass[sigconf, nonacm]{acmart}

\usepackage{float}
\newcommand\vldbdoi{XX.XX/XXX.XX}

\newcommand\vldbauthors{\authors}
\newcommand\vldbtitle{\shorttitle} 
\newcommand\vldbavailabilityurl{https://github.com/jam1729/DiskANN/tree/variable_quantization}
\newcommand\vldbpagestyle{plain} 
\usepackage{xcolor}
\usepackage{algorithm}
\usepackage{algpseudocode}
\usepackage{seqsplit}

\begin{document}
\title{Quantization Beyond Uniform Bit Allocation}

\author{K.~S.~Sreeramji}
\authornote{Work done during internship at Microsoft Research.}
\affiliation{%
	\institution{Indian Institute of Science}
	\city{Bengaluru}
	\state{India}
}
\email{sreeramjiks@iisc.ac.in}

\author{Sabyasachi Basu}
\affiliation{%
	\institution{Microsoft Research}
	\city{Bengaluru}
	\state{India}
}
\email{sabyasachi.basu@microsoft.com}

\author{Ravishankar Krishnaswamy}
\affiliation{%
	\institution{Microsoft Research}
	\city{Bengaluru}
	\state{India}
}
\email{rakri@microsoft.com}

\author{Kirankumar Shiragur}
\affiliation{%
	\institution{Microsoft Research}
	\city{Bengaluru}
	\state{India}
}
\email{kshiragur@microsoft.com}

\author{Yujia Wang}
\affiliation{%
	\institution{Microsoft STCA}
	\city{Suzhou}
	\state{China}
}
\email{yujiawang@microsoft.com}


\begin{abstract}
Quantization is a fundamental technique to handle the growing sizes of embeddings generated by modern models. Existing quantization schemes are largely embedding agnostic and allocate bits uniformly across dimensions. However, recent models produce embeddings with significant geometric structure. In this work, we investigate whether a variable bit allocation scheme can improve quantization quality under a fixed memory budget. We propose a simple variable bit allocation framework that partitions an embedding into contiguous buckets and allocates storage non-uniformly across them. Using a greedy allocation strategy, we instantiate this framework for both Product Quantization (PQ) and Scalar Quantization (SQ). We perform a series of experiments on embeddings known to have the Matryoshka property (MRL), and consistently observe that non-uniform allocations outperform uniform baselines at identical storage budgets. The largest improvements occur in the low-bit regime, where uniform allocation is particularly inefficient for MRL embeddings. At the same compression rates, variable allocation improves recall by up to 8\% for PQ and up to 18\% for SQ. Our results suggest a new direction for structure-aware compression and indexing techniques for large-scale retrieval systems.
\end{abstract}

\maketitle

\pagestyle{\vldbpagestyle}
\begingroup\small\noindent\raggedright\textbf{VLDB Workshop Reference Format:}\\
\vldbauthors. \vldbtitle.
VLDB 2026 Workshop: 2nd Workshop on Vector Databases (VecDB).
\\
\endgroup
\begingroup
\renewcommand\thefootnote{}\footnote{\noindent
	This work is licensed under the Creative Commons BY-NC-ND 4.0 International License.
	Visit \url{https://creativecommons.org/licenses/by-nc-nd/4.0/} to view a copy of this license.
	For any use beyond those covered by this license, obtain permission by emailing \href{mailto:info@vldb.org}{info@vldb.org}.
	Copyright is held by the owner/author(s).
	Publication rights licensed to the VLDB Endowment.
	\\
	\raggedright Proceedings of the VLDB Endowment.
	ISSN 2150-8097.
	\\
}\addtocounter{footnote}{-1}\endgroup

\ifdefempty{\vldbavailabilityurl}{}{
	\vspace{.3cm}
	\begingroup\small\noindent\raggedright\textbf{VLDB Workshop Artifact Availability:}\\
	The source code, data, and/or other artifacts have been made available at \url{\vldbavailabilityurl}.
	\endgroup
}

\section{Introduction}

Nearest neighbor search is a crucial component of nearly all forms of information retrieval in large ML systems that use embeddings: from search \cite{10.1145/3394486.3403305, LIU2007262}, to recommendations \cite{10.1145/2959100.2959190}.
However, due to the curse of dimensionality, nearest neighbor search becomes infeasible \cite{10.1145/276698.276876, 10.1145/1553374.1553485}; even the more popular approximate versions of nearest neighbor search used in practice \cite{diskann-github, scann_2020, soar_2023, 11202651, HNSW} face challenges related to scalability \cite{10.1145/3589282}. 
Modern embeddings routinely have thousands of dimensions; a single 3000 dimensional vector has a disk footprint of 12 kilobytes, implying a total footprint of several terabytes for billion scale datasets.
Quantization provides some reprieve in these situations: these techniques aim to reduce the storage and memory footprints by replacing each full vector with a quantized code that approximates the geometric structure of the dataset \cite{5432202, 11202651}.
Reducing the data type's footprint from a 32 bit float representation to a custom 1 bit binary representation reduces the footprint by a factor of 32, albeit at the cost of some loss in embedding quality.

Several quantization schemes have shown excellent performance on real-world data and are routinely used in large scale industrial systems.
Of these, Product Quantization (PQ)~\cite{5432202, pq-survey} and Scalar Quantization (SQ) \cite{11202651, scalar-1998, scalar-2023-blink} are perhaps the most well-known and widely used.

Methods such as this stand out for not just their impressive performance, but also their relative conceptual simplicity.

\begin{figure*}
	\centering
	\includegraphics[width=\textwidth]{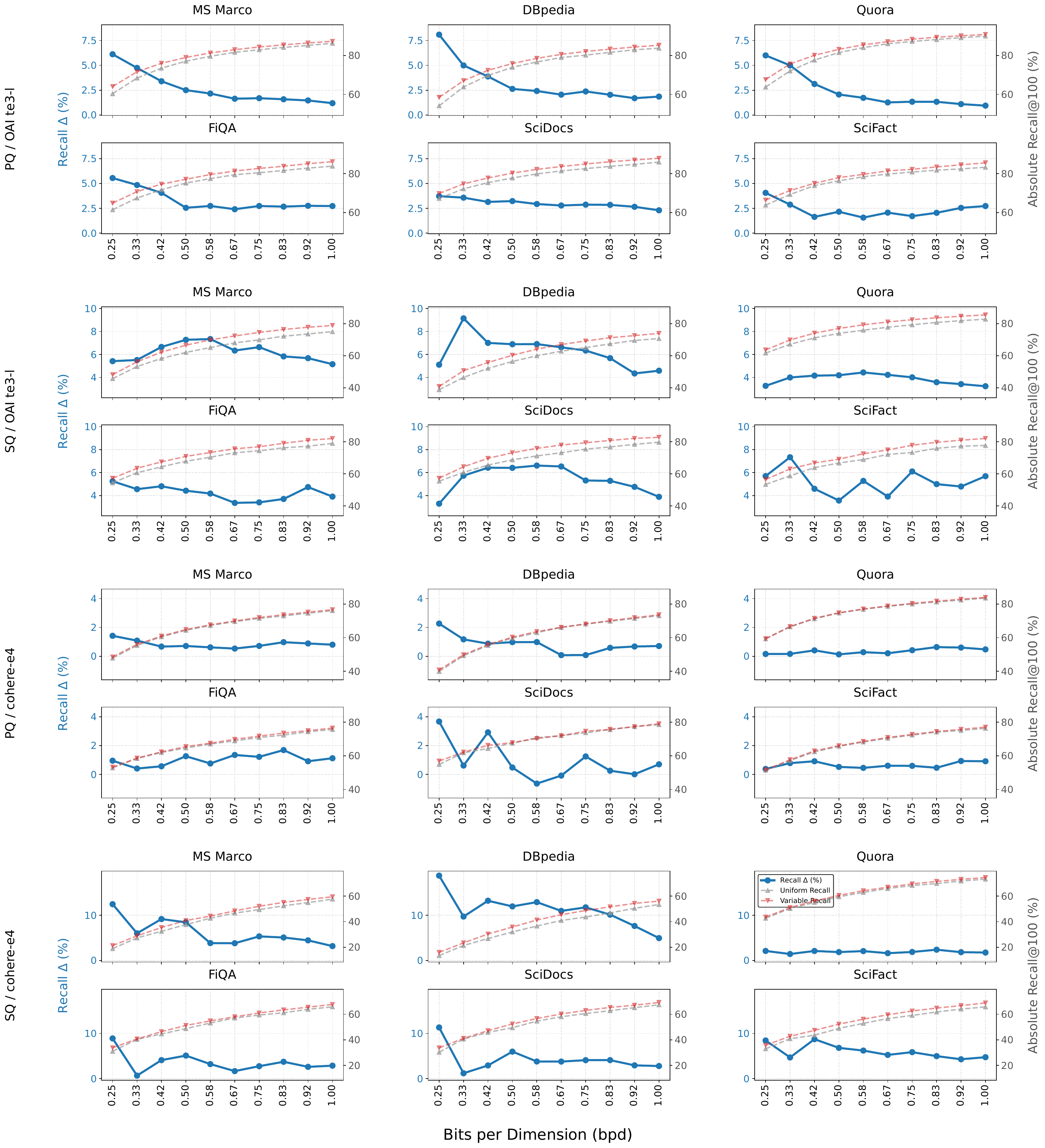}
	\caption{A comprehensive account of the gains in recall across different datasets.
		We compare gains on embeddings generated using OpenAI's \texttt{{text-embedding-3-large}} and Cohere's \texttt{{embed-v4}} models, using both Product Quantization (PQ) and Scalar Quantization (SQ).
		In addition to the relative recall improvement, we also present the absolute 100-recall@100 achieved by the uniform and variable allocation schemes.
	}
	\label{fig:main-gains}
\end{figure*}
Furthermore, many modern embeddings \cite{openai2024embeddings, shanbhogue2026geminiembedding2native, vera2025embeddinggemmapowerfullightweighttext, zhang2025qwen3embeddingadvancingtext, jina-v5-text, voyageai2026voyage4, nomic_embed_v1_5_hf, cohere_embed_v4_release} are known to exhibit the \emph{Matryoshka property} (MRL) \cite{KBR+22}, where the leading dimensions capture a disproportionate fraction of the structure of the embeddings.
This implies that even if the full vector itself is quite large, truncating it to a fraction of its length and retaining only the first few dimensions can preserve the vector's quality sufficiently well for most downstream tasks, including retrieval.
In fact, prior work observes that across many real-world datasets, retaining just the top 15\% of dimensions achieves over 70\% recall from modern embeddings that have the MRL property (Figure 4 of \cite{KBR+22}).

So far, post training quantization schemes have been largely embedding agnostic.
We ask the following question at this juncture:
\begin{quote}
	Can a variable bit allocation scheme allow for better quantization of vectors at the same compression rate?
\end{quote}
Specifically, in the setting of MRL embeddings, this manifests as an interesting hypothesis: given the outsized importance of the initial dimensions, a strategy that assigns more bits to the initial dimensions may outperform existing techniques that provide uniform (or amortized uniform) bit budgets to all dimensions.

This paper affirms this hypothesis.
We show that variable allocation strategies allow us to beat standard uniform allocation quantization schemes at the same bit budgets.  All embeddings used in our experiments are known to have the MRL property (refer to Figure~\ref{fig:variance_plot_128}). While our allocation scheme does not explicitly make use of the Matryoshka property, the ``optimized'' bit allocation scheme strongly reflects the property, with more bits being assigned to the leading dimensions. 
Specifically, at low bit budgets, we observe that aggressively assigning more bits to the leading dimensions leads to noticeable gains over uniform allocation for the same quantization techniques.
We evaluate Scalar and Product Quantization (hereafter referred to as SQ and PQ respectively) using a greedy bit allocation scheme with fixed buckets that group consecutive dimensions.
Under this scheme, the algorithm is granted a bit budget in fixed increments, and assigns the additional bits to the bucket which gives the greatest improvement.
Under this scheme, we see that there is up to an 8\% relative improvement over uniform for PQ and over 18\% for SQ when measuring 100-recall@100.
The gains are asymmetric and both model and dataset dependent; however, even if margins are thin, there are consistent gains. 
An important feature of this approach is the improvement in the sub-1 bit regime. Here, uniform bit allocation is demonstrably suboptimal for vectors with the MRL property (refer to Figure~\ref{fig:main-gains}).
Unsurprisingly, our biggest wins are consistently in this regime.

This work adopts an exploratory, data-driven approach to this problem.
However, the greedy allocation scheme as described here is inefficient and especially slow on large datasets.
An interesting future direction would be to design a fast algorithm to determine optimal allocation, and explore a closed form expression for bit allocation under the Matryoshka property.

\section{Related Work}

Vector quantization finds its roots in Shannon's work in compression and rate distortion functions~\cite{Shannon59, Huang1963BlockQO, gersho1992vector}.
The simplest of these is Scalar Quantization (SQ) \cite{11202651, scalar-1998, scalar-2023-blink}, which involves discretizing the range of values in a representation uniformly to fewer bits of precision.
In the context of embeddings, Product Quantization (PQ) \cite{5432202, pq-survey} has found ubiquity in reducing the sizes of indices for nearest neighbor search.
PQ is used very commonly in industry and is integrated into several nearest-neighbor libraries as a default.
While PQ uses $k$-means on a group of dimensions (commonly referred to as ``subvectors''/``chunks''), other techniques inspired by it have also tried out other clustering variants.
Recent work has also adapted PQ for online settings by removing the need for pre-processing and look up tables, and modern SIMD-based implementations are incredibly fast \cite{pq-simd, pq-simd-2}.
However, typically, these techniques do not provide any guarantees on the quality of the quantized vector in terms of error bounds.
More recently, RaBitQ \cite{rabitq} provided a quantization scheme that produces a bit string and provides an unbiased distance estimator with provable, asymptotically optimal error bounds.
Moreover, RaBitQ and its variants show superior performance over PQ-style quantizers.

Matryoshka Representation Learning (MRL) was first proposed by Kusupati et al. \cite{KBR+22},  to allow a single representation to capture information at multiple granularities.
Specifically, vectors exhibiting this property allow users to drastically reduce the footprint and computational overhead of using the full vector, with minimal preprocessing (such as truncation).
This allows for the same embedding to be used in multiple downstream tasks with varying computational capacities with virtually no additional cost.
In the years since its introduction, multiple large embedding models have defaulted to producing vectors that exhibit the Matryoshka property.

There has been recent work on dynamic quantization schemes \cite{chen2026annsampacceleratingapproximatenearest, tewary2026aqrhnswacceleratingapproximatenearest} that adaptively adjust precision at query time under latency constraints.
However, they are embedding-agnostic and impose significant computational overhead.
Our approach, instead, leverages the inherent structural properties of the embeddings.
The adaptivity in our scenario occurs offline during index construction.
We use the inherent variance decay in MRL embeddings to perform offline bit allocation.
While our work does not focus on an efficient implementation, we still maintain query time performance of static, uniform-width codebooks. There is also a growing body of work on quantization aware training ~\cite{JKC+18, Zhang2025SurveyOQ}; these methods simulate low-precision arithmetic during the training process with great success. However, they incur significant training overhead. Our approach, in contrast, is post-training quantization, which is more commonly used in practice for its ease in deployment and lower cost.

\section{A Greedy Bit Allocation Scheme}

The core idea behind quantization is that lower precision in data is often sufficient to maintain the global structure of the data.
This indicates that for most practical purposes, one can reduce the range of values required to represent the data.
Therefore, unlike the traditional 32-bit float data type that is typically used by default, one can reduce the bit complexity and use data types with a smaller bit footprint.

\subsection{Matryoshka embeddings}
Our work focuses on embeddings generated by \emph{Matryoshka Representation Learning} (MRL) training, a line of investigation initiated by Kusupati et al.~\cite{KBR+22}.
Suppose we have a labeled dataset $\mathcal{D} = \{(x_1,y_1), \dots, (x_N,y_N)\}$, where $x_i \in \mathcal{X}$ and $y_i \in [L]$.
MRL optimizes a multi-class softmax cross-entropy loss $\mathcal{L}: \mathbb{R}^L \times [L] \to \mathbb{R}_+$ across a set of nested dimensions $m \in \mathcal{M}$.

Each nested dimension $m$ utilizes a separate linear classifier parameterized by $\mathbf{W}^{(m)} \in \mathbb{R}^{L \times m}$.
The overall objective minimizes the weighted empirical risk over the feature extractor parameters $\theta_F$ and classifiers:

\begin{equation}
	\min_{\{\mathbf{W}^{(m)}\}_{m \in \mathcal{M}}, \, \theta_F} \frac{1}{N} \sum_{i \in [N]} \sum_{m \in \mathcal{M}} c_m \cdot \mathcal{L}\left(\mathbf{W}^{(m)} \cdot F(x_i; \theta_F)_{1:m} \;;\; y_i\right).
\end{equation}

Here $F(x_i; \theta_F)_{1:m}$ denotes the truncation of the extracted feature to its first $m$ coordinates. $c_m \geq 0$ denotes the relative importance scale of dimension $m$ (set to $c_m = 1, \forall m \in \mathcal{M}$ by default).

While the MRL property is inherent to the loss function, a strict mathematical or geometric definition for the resulting embeddings remains elusive.
Through empirical investigation, however, we identify an exploitable geometric proxy: the decay in variance across dimensions as the dimension index increases.
By construction, the nested objective concentrates information in the initial coordinates, resulting in higher variance in the early dimensions and lower residual variance in the later dimensions.

\begin{figure}[H]
	\centering
	\includegraphics[width=\linewidth]{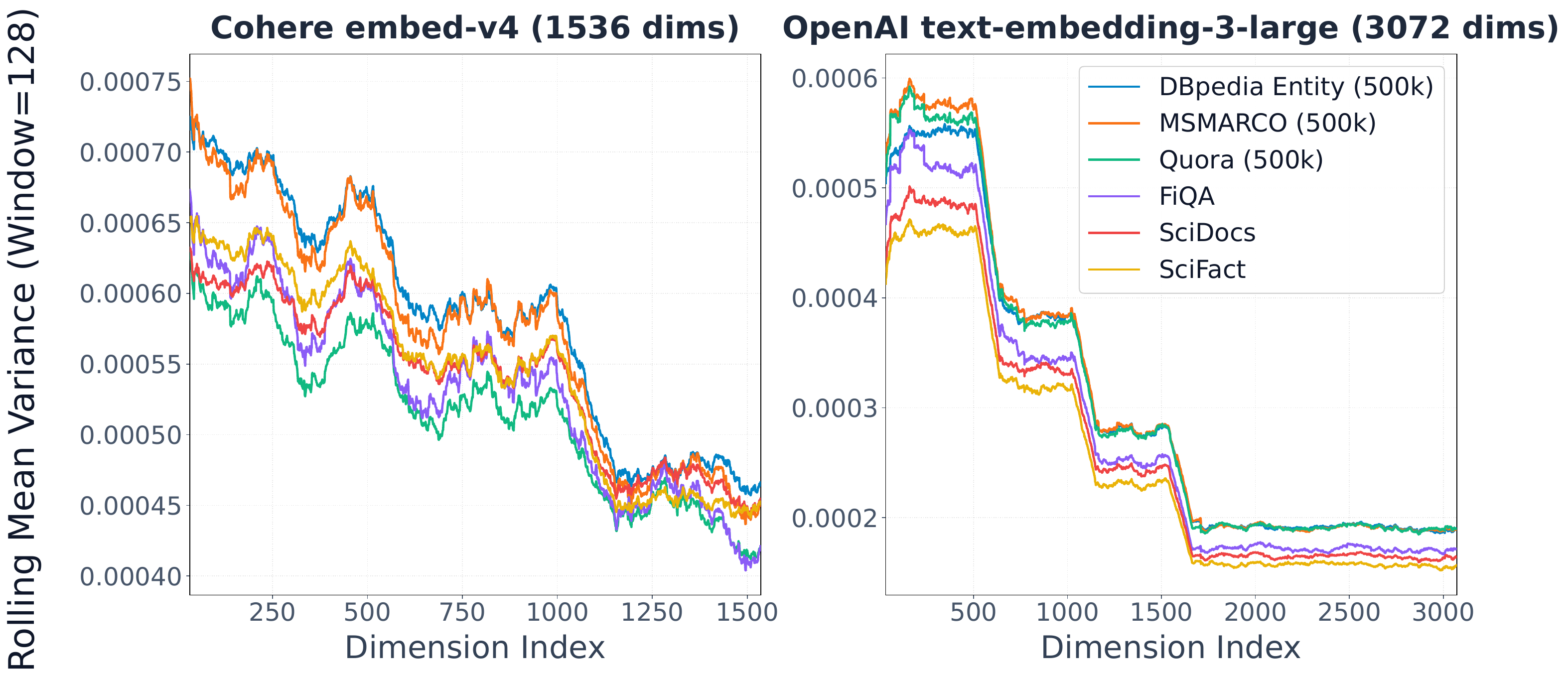}
	\caption{Observed variance differences by dimension in embeddings produced by two different embedding models: OpenAI's \texttt{text-embedding-3-large} and Cohere's \texttt{embed-v4}.
		We plot the rolling window average variance per dimension across different datasets and observe very prominent iso-variance ``levels'' across the dimensions. The rolling window is 128 dimensions in length.
	}
	\label{fig:variance_plot_128}
\end{figure}

This hierarchy of variances manifests in embeddings from models trained with MRL; Figure~\ref{fig:variance_plot_128} illustrates this decay in recent models from both Cohere and OpenAI.
This variance decay provides an intuitive explanation for why Matryoshka embeddings succeed in practice.
When all dimensions of an embedding exhibit similar levels of variance, they carry equal importance. However, concentrating variance in the initial dimensions mimics the leading digits of a numerical value, creating a standalone, coarser (but still informative) representation without the trailing dimensions.

\subsection{Greedy Allocation Framework}
To systematically evaluate the paper's hypothesis, we propose a variable-bit allocation scheme that allows room for exploiting geometric structure in the embeddings. The framework partitions consecutive dimensions into contiguous buckets and assigns varying bit budgets to each, naturally leading to a structure-aware, variable-bit quantization scheme.

We partition the total embedding dimensionality $D$ into $K$ contiguous buckets $B_1, B_2, \dots, B_K$. 
The size of bucket $B_k$ is represented as $d_k = |B_k|$, such that $\sum_{k=1}^K d_k = D$.
Instead of starting the search from zero bits—which would trivially yield zero recall—the search initializes with a uniform byte allocation vector $\vec{b}^{(0)} = (b_{\text{init}}, \dots, b_{\text{init}})$ across all buckets.

We perform a greedy local search to find an optimized bit allocation vector $\vec{b} = (b_1, \dots, b_K)$, where $b_k \ge 0$ represents the budget (in bytes) allocated to bucket $B_k$.
At each iteration, the algorithm evaluates the marginal utility of allocating more memory to a specific region of the embedding. We consider candidate configurations by temporarily increasing the bit budget of each bucket in turn by a fixed step-size increment $\delta$ (in bytes). 
We then evaluate the search recall $R(\vec{b}, \mathcal{V})$ on a validation query set $\mathcal{V}$ using exact distance computation over the quantized dataset.
The candidate bucket that yields the maximum empirical recall "wins" the iteration and permanently receives the $\delta$ bits.
The search proceeds iteratively until the predefined global target memory budget is achieved.
This procedure is detailed in Algorithm~\ref{alg:greedy_search}.
Note that our algorithm does not explicitly focus on MRL embeddings. However, our results show that this scheme heavily concentrates bits in the leading dimensions, thereby recovering the latent MRL structure that we know these embeddings possess.

\begin{algorithm}[H]
	\caption{Greedy Bit Allocation}\label{alg:greedy_search}
	\begin{algorithmic}[1]
		\Statex \textbf{Input:}
		Initial uniform allocation $\vec{b}^{(0)} = (b_{\text{init}}, \dots, b_{\text{init}})$, step-size increment $\delta$, validation set $\mathcal{V}$ \Statex \textbf{Output:} Optimized allocation vector $\vec{b}$ \State $\vec{b} \gets \vec{b}^{(0)}$ \Repeat \State $k^* \gets \arg\max_{k \in \{1, \dots, K\}} R(\vec{b} + \delta \vec{e}_k, \mathcal{V})$ 
        \State $\vec{b} \gets \vec{b} + \delta \vec{e}_{k^*}$
        \Until{target budget is achieved} \State \Return $\vec{b}$ \end{algorithmic} \end{algorithm}

Note that the algorithm's runtime is linearly proportional to $K$ times the number of increments needed to reach the target budget.

\subsubsection{Greedy Product Quantization} 
In standard Product Quantization (PQ) \cite{5432202}, a high-dimensional vector space is decomposed into lower-dimensional, mutually orthogonal subspaces. These subspaces are quantized independently using separate $k$-means codebooks, allowing a massive number of effective centroids to be represented compactly. In our greedy variable-bit paradigm, we apply PQ locally within each bucket $B_k$, strictly constrained by its dynamically assigned byte budget. 

Specifically, when a bucket $B_k$ of size $d_k$ receives a budget of $b_k$ bytes, we must partition its $d_k$ dimensions into exactly $b_k$ contiguous subvectors. By utilizing codebooks with $C = 256$ centers, the integer index of the nearest centroid for each subvector fits perfectly into exactly $\log_2(256) = 8$ bits ($1$ byte) of storage. Thus, the total byte budget precisely dictates the number of subvectors a bucket is split into.

Because $d_k$ is rarely perfectly divisible by $b_k$ during a dynamic allocation, we handle dimension remainders by front-loading them. The remainder dimensions $r = d_k \bmod b_k$ are distributed evenly by adding exactly one extra dimension to the first $r$ subvectors. For example, if a bucket of $10$ dimensions is allocated a $3$-byte budget, it is partitioned into three subvectors of lengths $4, 3$, and $3$.
This dynamic subvector partitioning and quantization procedure is outlined in Algorithm~\ref{alg:pq_subroutine}\footnote{In every iteration of the greedy search, the algorithm evaluates $K$ candidate allocations. For PQ, evaluating a candidate requires training new $k$-means codebooks for the upgraded bucket. To keep offline index construction tractable, the trained codebooks $\mathcal{C}_{k}(b)$ are cached for each bucket $k$ and byte-budget $b$, ensuring the clustering algorithm is executed only once per configuration.}.

\begin{algorithm}[H]
	\caption{PQ Dimension Partitioning}\label{alg:pq_subroutine}
	\begin{algorithmic}[1]
		\Statex \textbf{Input:} Sub-vector $x^{(k)} \in \mathbb{R}^{d_k}$ of bucket $B_k$, budget $b_k$ (in bytes), bucket training data $\mathcal{X}_{\text{train}}^{(k)}$
		\Statex \textbf{Output:} Quantized code vector $\vec{q}^{(k)} \in \{0, \dots, 255\}^{b_k}$
		\State $s \gets \lfloor d_k / b_k \rfloor$, $r \gets d_k \bmod b_k$
		\State Partition $x^{(k)}$ and $\mathcal{X}_{\text{train}}^{(k)}$ into $b_k$ contiguous subvectors, where the first $r$ subvectors have dimension $s+1$ and the remaining have dimension $s$.
		\For{each subvector index $j = 1$ \textbf{to} $b_k$}
			\State Train codebook $\mathcal{C}_{k, j}$ with 256 centers using $k$-means
			\Statex \hskip\algorithmicindent on the $j$-th partition of $\mathcal{X}_{\text{train}}^{(k)}$
			\State Quantize $x^{(k)}_j$ to closest centroid:
			\Statex \hskip\algorithmicindent $\vec{q}^{(k)}[j] \gets \arg\min_{c \in \mathcal{C}_{k, j}} \| x^{(k)}_j - c \|_2$
		\EndFor
		\State \Return $\vec{q}^{(k)}$
	\end{algorithmic}
\end{algorithm}
\begin{figure*}
	\centering
	\includegraphics[width=\textwidth]{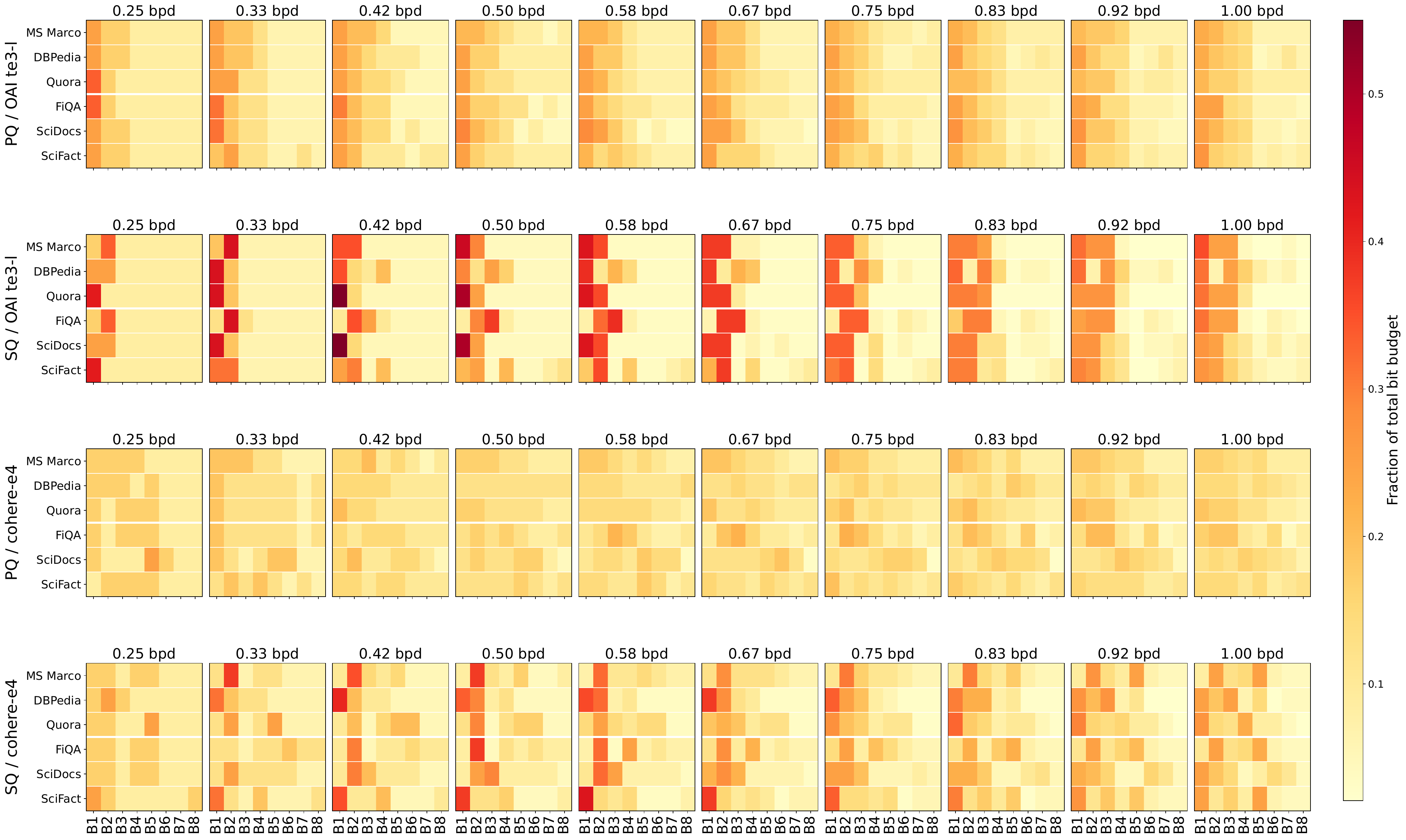}
	\caption{Optimized bit allocation across buckets at different bit budgets, noted by bits per dimension (bpd). Observe that optimized allocation concentrates bits towards the initial buckets, in line with what we expect due to the MRL property.
	}
	\label{fig:bit-alloc}
\end{figure*}
\subsubsection{Greedy Scalar Quantization} 
Scalar Quantization (SQ) \cite{11202651} compresses embeddings by independently discretizing the continuous coordinate values of each dimension into a finite set of representation levels. In our variable-bit paradigm, we represent each dimension using a code with a specific bit-width $w_i$ selected from $\{0, 2, 4, 8\}$\footnote{A $1$-bit quantization splits the distribution directly at its mode, artificially pushing high-density central values to the extremes. To preserve the unimodal nature of standard embedding coordinates, $1$-bit widths are explicitly avoided. A bit-width of $0$ corresponds to completely discarding, or pruning, the dimension.}.

A budget of $b_k$ bytes allocates $N_{\text{bits}} = 8 b_k$ total bits to the bucket of size $d_k$. While standard SQ operates comfortably at higher, uniform bit-widths, modern scale often pushes systems into the extreme sub-$1$ bit or sub-$2$ bit regime. In these constrained environments, it is impossible to allocate even the minimum $2$-bit width to every dimension. The algorithm must selectively drop (assign $0$ bits to) a fraction of the coordinates. 

To maintain structural integrity without introducing complex indexing metadata, we distribute these sparse bits as evenly as possible across the $d_k$ dimensions using a uniform stride allocation. We begin by establishing a baseline bit-width $W_{\text{base}} \in \{0, 2, 4\}$ across all dimensions in the bucket. To allocate any remaining bits, we select $u$ dimensions to upgrade to the next higher level $W_{\text{next}} \in \{2, 4, 8\}$ by stepping through the bucket with a uniform stride $\Delta = \lfloor d_k / u \rfloor$. For instance, a $192$-dimension bucket with a bit budget $N_{\text{bits}} = 192$ has a baseline of $0$ bits and receives $96$ $2$-bit upgrades -- we assign $2$ bits to every second dimension ($\Delta = 2$) and discard the other dimensions.

Once the bit-widths $\{w_i\}$ are assigned, each dimension is quantized using standard SQ at its respective bit-width. 
This deterministic allocation and scaling procedure is detailed in Algorithm~\ref{alg:sq_subroutine}.

\begin{algorithm}[H]
	\caption{SQ Bit-Width Allocation}\label{alg:sq_subroutine}
	\begin{algorithmic}[1]
		\Statex \textbf{Input:} Sub-vector $x^{(k)} \in \mathbb{R}^{d_k}$ of bucket $B_k$, budget $b_k$ (in bytes)
		\Statex \textbf{Output:} Quantized code vector $\vec{q}^{(k)}$
		\State Define allowed bit-widths: $S \gets \{0, 2, 4, 8\}$
		\State $N_{\text{bits}} \gets 8 \cdot b_k$
		\State $W_{\text{base}} \gets \max \{ w \in S \mid w \cdot d_k \le N_{\text{bits}} \}$
		\If{$W_{\text{base}} = 8$}
			\State $W_{\text{next}} \gets 8$
			\State $u \gets 0$
		\Else
			\State $W_{\text{next}} \gets \min \{ w \in S \mid w > W_{\text{base}} \}$
			\State $u \gets (N_{\text{bits}} - d_k \cdot W_{\text{base}}) / (W_{\text{next}} - W_{\text{base}})$
		\EndIf
		\State $w_{i} \gets W_{\text{base}}$ \quad for $i = 1, \dots, d_k$
		\If{$u > 0$}
			\State $\Delta \gets d_k / u$
			\State Upgrade dimensions along the stride: 
			\Statex \hskip\algorithmicindent $w_{\text{round}(j \cdot \Delta) + 1} \gets W_{\text{next}}$ \quad for $j = 0, \dots, u-1$
		\EndIf
		\For{each dimension $i = 1$ \textbf{to} $d_k$}
			\State $\vec{q}^{(k)}[i] \gets \text{ScalarQuantize}(x^{(k)}_i, w_i)$
		\EndFor
		\State \Return $\vec{q}^{(k)}$
	\end{algorithmic}
\end{algorithm}

\section{Experiments} \label{sec:experiments}

\paragraph{System} All experiments were run on a Linux x86\_64 machine with 16 hardware threads on an Intel Xeon Platinum 8272CL CPU at 2.60\,GHz and 32\,GiB of main memory, with no GPU.
We built DiskANN to use its quantization utilities, using the repository's standard CMake/ C++17 toolchain and the documented Linux dependencies for OpenMP, Boost, and Intel MKL. We do not use DiskANN's graph indexing. We use brute-force L2 distance computation to compute exact Nearest Neighbors at all instances. The code is branched from Microsoft's DiskANN repository in C++ and a link to our implementation is provided in the list of artifacts.

\paragraph{Datasets.}
We evaluate our results on MS Marco \cite{DBLP:conf/nips/NguyenRSGTMD16}, DBpedia-Entity \cite{Hasibi:2017:DVT}, FiQA \cite{fiqa}, SciDocs \cite{scidocs}, SciFact \cite{scifact} and Quora; some of the standard datasets for evaluating quantization techniques (MTEB \cite{muennighoff2022mteb,enevoldsen2025mmtebmassivemultilingualtext}).
We draw these datasets from the BEIR benchmark~\cite{thakur2021beir} to ensure standardization and reproducibility.
For the larger datasets (MS Marco, DBpedia, Quora), the base corpus is truncated to the first 500,000 entries. The smaller datasets are used in their entirety.
The evaluation queries are taken from the standard evaluation split, as detailed in Table ~\ref{tab:dataset-stats}.

\begin{table}[h]
\centering
\begin{tabular}{lrrr}
\toprule
Dataset & Base Vectors & Queries & Split \\
\midrule
MS Marco & 500,000 & 6,980 & Dev \\
DBpedia-Entity & 500,000 & 400 & Test \\
Quora & 500,000 & 10,000 & Test \\
FiQA & 57,638 & 648 & Test \\
SciDocs & 25,657 & 1,000 & Test \\
SciFact & 5,183 & 300 & Test \\
\bottomrule
\end{tabular}
\caption{Datasets.}
\label{tab:dataset-stats}
\end{table}

Quora represents a symmetric duplicate question detection task (where queries and corpus entries are short questions of similar length), complementing the asymmetric retrieval characteristics of MS Marco and DBpedia-Entity. 
Furthermore, we also include results on some smaller datasets: FiQA, SciDocs, and SciFact. Structurally, they offer a mix of retrieval paradigms: FiQA and SciFact feature asymmetric tasks mapping short questions or claims to longer text paragraphs, whereas SciDocs evaluates document-to-document retrieval where both queries and corpus entries are of similar length.

\paragraph{Embedding models.}
We use one of OpenAI's most capable commercial embedding models, \texttt{\seqsplit{text-embedding-3-large}} \cite{openai2024embeddings}, and Cohere's latest embedding model, \texttt{\seqsplit{embed-v4}} \cite{cohere_embed_v4_release}.
The full dimensionality of \texttt{\seqsplit{text-embedding-3-large}} is $3072$, and that of \texttt{\seqsplit{embed-v4}} is $1536$.
We generate full-dimension embeddings from both models.

\paragraph{Metric.}
We compute the average $100$-recall@$100$ over the query set.
Crucially, the recall is measured via brute-force exact L2 (Euclidean) distance search over the dequantized (inflated) database vectors.
This isolates the quantization error itself, ensuring the metric is independent of any error introduced by approximate indexing schemes.

\paragraph{Setup.}
The baseline uses uniform allocation of bits across all dimensions. To facilitate a direct comparison between embeddings of different dimensionalities, we present our results in terms of Bits per Dimension (bpd).
We evaluate budgets starting from $\approx 0.17$ bits per dimension ($1/6$ bpd) up to 1 bit per dimension, with a step size of $1/12$ bits per dimension.
For greedy allocation, we split the full dimensions into 8 equal contiguous buckets.
Each bucket has 384 dimensions for \texttt{\seqsplit{text-embedding-3-large}} ($3072 / 8 = 384$) and 192 dimensions for \texttt{\seqsplit{embed-v4}} ($1536 / 8 = 192$).

\paragraph{Hyperparameters.}
The greedy search starts with an initial uniform allocation of 64 bytes (8 bytes for each of the 8 buckets) for embeddings from \texttt{\seqsplit{text-embedding-3-large}} and of 32 bytes (4 bytes for each of the 8 buckets) for embeddings from \texttt{\seqsplit{embed-v4}}.
At each iteration, we consider candidate configurations by increasing the bit budget of each bucket in turn by a step-size increment of $\delta = 8$ bytes and $\delta = 4$ bytes respectively for \texttt{\seqsplit{text-embedding-3-large}} and \texttt{\seqsplit{embed-v4}}.
The sweep runs for 40 iterations, at which point it reaches the target total budget of 1 bit per dimension, which is 384 bytes and 192 bytes for \texttt{\seqsplit{text-embedding-3-large}} and \texttt{\seqsplit{embed-v4}} respectively.
During the training phase (e.g. for computing product quantization codebooks or scalar quantization scaling factors), we random-sample exactly 10\% of the base corpus vectors.

\paragraph{Observations}
Figure~\ref{fig:main-gains} shows the gains over uniform achieved by our greedy allocation scheme. 
Moreover, the winning bit allocations assign more bits to the initial dimensions, as noted in Figure~\ref{fig:bit-alloc}. The asymmetry of the optimal bit allocations discovered by our greedy algorithm varies across settings; overall, it is more pronounced in scalar quantizers, and this coincides with the instances of highest relative gains. On the other hand, the PQ case exhibits smoother decay in bit allocation from the leading to the trailing dimensions.
For completeness, we also compare with straightforward truncation to the leading dimensions according to the bit budgets (see Table~\ref{tab:truncation-recall}). 
While Figure~\ref{fig:main-gains} focuses on the improvement over the uniform baseline, it also plots the actual recall achieved in both the uniform and variable cases. Note the diminishing returns at higher bit budgets, which correspond to the already high baseline recall. 
We observe that the gains from a variable bit allocation scheme are typically higher for the OpenAI embeddings; this is in line with our observation in Figure~\ref{fig:variance_plot_128}, where we notice that the variance levels are more distinct in case of the embeddings generated by  \texttt{\seqsplit{text-embedding-3-large}}.

\section{Limitations and Future Work}
We present this work as a preliminary, exploratory exposition on variable allocation schemes for Matryoshka embeddings. Our results tell us that it is \emph{possible} to improve over the uniform bit allocation process using a variable scheme. However, several important questions remain, we list some concrete directions below:
\begin{enumerate}
    \item \emph{A concrete mathematical description of MRL embeddings: } The variance observations in Figure~\ref{fig:variance_plot_128} show that the variance is a good proxy for the MRL property. However, it remains to show whether a concrete mathematical or geometric description of the embeddings can be derived from the loss function. This would also help formalize a quantization scheme that explicitly exploits the MRL property.
    \item \emph{An efficient allocation algorithm: } The current greedy approach is extremely expensive as it iterates over a lot of candidate allocations. Moreover, this is not a practical approach to finding the best scheme. Moreover, our experiments are limited to fixed, discrete buckets; it is possible that a better quantization exists by taking finer buckets; in our greedy approach, this would blow up the complexity of finding the optimized scheme even more. A practical quantizer requires a more efficient implementation, potentially leveraging a deterministic heuristic, or a lightweight randomized/iterative process. 
    \item \emph{Systems challenges: } Dynamic bit allocation introduces significant systems challenges. Storing large volumes of data using custom-sized data types requires adapting existing architectures. This sacrifices the memory access efficiency provided by uniform data types, posing a particular challenge for SIMD-friendly architectures. Especially with constraints like latency, this poses a challenging problem. 
    \item \emph{Ablation studies: } A thorough, ablation study of the setup with different numbers of buckets, further metrics (such as recall at different thresholds), and further hyperparameter tuning across further models is required for a better understanding of this landscape. 
\end{enumerate}
\begin{table}[t]
\centering

\setlength{\tabcolsep}{4pt}
\small
\begin{tabular}{lcccccc}
\toprule
\multicolumn{7}{l}{\textit{Cohere embed-v4}} \\
\midrule
& \multicolumn{6}{c}{Bits per Dimension} \\
\cmidrule(lr){2-7}
Dataset & 0.17 & 0.33 & 0.50 & 0.67 & 0.83 & 1.00 \\
        & (32\,B) & (64\,B) & (96\,B) & (128\,B) & (160\,B) & (192\,B) \\
        & (8d)  & (16d) & (24d) & (32d)  & (40d)  & (48d)  \\
\midrule
MS\,Marco  & 0.24 & 1.04 & 2.30 & 3.87 & 5.92 &  8.64 \\
DBpedia    & 0.20 & 0.68 & 1.54 & 2.38 & 3.76 &  5.47 \\
Quora      & 1.95 & 7.13 & 12.59 & 17.46 & 22.62 & 27.62 \\
FiQA       & 1.19 & 2.98 &  5.52 &  7.72 &  9.70 & 12.61 \\
SciFact    & 5.73 & 8.22 & 10.66 & 12.66 & 15.36 & 17.79 \\
SciDocs    & 2.08 & 4.15 &  6.75 &  8.76 & 10.61 & 12.93 \\
\midrule
\multicolumn{7}{l}{\textit{OpenAI text-embedding-3-large}} \\
\midrule
& \multicolumn{6}{c}{Bits per Dimension} \\
\cmidrule(lr){2-7}
Dataset & 0.17 & 0.33 & 0.50 & 0.67 & 0.83 & 1.00 \\
        & (64\,B) & (128\,B) & (192\,B) & (256\,B) & (320\,B) & (384\,B) \\
        & (16d) & (32d)  & (48d)  & (64d)  & (80d)  & (96d)  \\
\midrule
MS\,Marco  & 1.80 &  6.65 & 13.53 & 19.98 & 25.97 & 31.38 \\
DBpedia    & 1.21 &  4.60 &  9.63 & 14.86 & 20.02 & 24.73 \\
Quora      & 7.18 & 19.32 & 31.13 & 39.85 & 46.05 & 50.97 \\
FiQA       & 5.67 & 13.59 & 21.34 & 27.93 & 34.02 & 38.96 \\
SciFact    & 10.68 & 17.93 & 24.52 & 29.58 & 34.02 & 38.19 \\
SciDocs    & 5.84 & 12.71 & 20.77 & 27.41 & 32.96 & 37.63 \\
\bottomrule
\end{tabular}
\caption{Recall@100 for simple dimension truncation (no quantization)
         across different compression rates (bits per dimension). 
         The absolute memory limits (in bytes) and their corresponding 
         32-bit float truncation dimensions are shown in parentheses.}
\label{tab:truncation-recall}
\end{table}

\begin{acks}
 K. S. Sreeramji was supported by the the Walmart Centre for Tech Excellence at IISc.
\end{acks}

\bibliographystyle{ACM-Reference-Format}
\bibliography{sample}

\end{document}